\documentclass[%
reprint,
superscriptaddress,
nofootinbib,
showkeys,
amsmath,amssymb,
aps,
pra,
longbibliography,
nobalancelastpage
]{revtex4-2}

\usepackage{common}

\begin{document}


\title{General Time-Dependent Configuration-Interaction Singles\\ \generalcase: The Molecular Case}

\author{Stefanos Carlström\,\orcidlink{0000-0002-1230-4496}}%
\email{stefanos@mbi-berlin.de}
\email{stefanos.carlstrom@matfys.lth.se}
\affiliation{Max-Born-Institut, Max-Born-Straße 2A, 12489 Berlin, Germany}
\affiliation{Department of Physics, Lund University, Box 118, SE-221 00 Lund, Sweden}

\author{Michael Spanner\,\orcidlink{0000-0003-3565-8518}}
\affiliation{National Research Council Canada, 100 Sussex Dr., Ottawa, Ontario K1A 0R6, Canada}

\author{Serguei Patchkovskii}%
\affiliation{Max-Born-Institut, Max-Born-Straße 2A, 12489 Berlin, Germany}

\date{\today}

\begin{abstract}
  We present a grid-based implementation of the time-dependent
  configuration-interaction singles method suitable for computing the
  strong-field ionization of small gas-phase molecules. After
  outlining the general equations of motion used in our treatment of
  this method, we present example calculations of strong-field
  ionization of \helium{}, \lih{}, \water{}, and \ethylene{} that
  demonstrate the utility of our implementation. The following
  companion paper \cite{Carlstroem2022tdcisII} specializes to the case
  of spherical symmetry, which is applied to various atoms.
\end{abstract}

\maketitle

\section{Introduction}
Strong-field physics and attosecond science offer a rich experimental
platform to study ultrafast phenomena on electronic time and length
scales \cite{Krausz2009,Scrinzi2006JoPBAMaOP}.  Applied to gas phase
molecules, techniques such as high-harmonic spectroscopy
\cite{Smirnova2009,Soifer2010,Worner2010,Worner2011,Kneller2022},
orbital tomography \cite{Itatani2004}, laser-induced electron
diffraction \cite{Spanner2004,Yurchenko2004,Meckel2008} and holography
\cite{Huismans2010}, attempt to use the technologies of attosecond
physics to probe electronic structures and dynamics that occur within
molecular systems.  Although promising, fine-tuning these techniques
to extract accurate and complete information is not straight-forward
due to the highly non-linear and non-perturbative nature of the
strong-field interactions driving these experimental efforts. Often
the only path to disentangling and interpreting the measurable
experimental observables is through detailed modeling of not only the
underlying molecular and electronic structures but also the complete
probing process itself. In this regard, the novel spectroscopic
methods brought forth by strong-field and attosecond physics will only
be as accurate as the underlying modeling used to interpreting the
complex observables involved.

The first step in all of these strong-field driven processes is the
removal of an electron with a strong low-frequency laser field.
Following ionization, the liberated electron is then accelerated in
the laser field and driven back to recollide with the parent ion, a
process called laser-induced electron recollision.  In order to
accurately describe this time-dependent non-perturbative process at an
\emph{ab initio} level, it is necessary to develop time-domain methods
that can interface with standard methods in time-independent
electronic structure theory.  In addition, a proper description of the
recollision motion of the continuum electron will likely require going
beyond the localized Gaussian-like basis sets upon which most of
standard electronic structure codes are based.  In this article, we
develop and explore a grid-based implementation of Time-Dependent
Configuration-Interaction Singles (TD-CIS) as applied to small
gas-phase molecules, where Cartesian grids are used to represent the
continuum electron, while still using Gaussians for the occupied
orbitals.

The problems we are interested in studying are not new, and many
different approaches have been fruitfully pursued before. A listing,
that is by no means exhaustive, would include TD-CIS for molecules
\cite{Klamroth2003,Krause2005,Schlegel2007,Krause2014,Krause2014a,Krause2015,Saalfrank2020},
TD-CIS for atoms \cite{Rohringer2006,Greenman2010PRA}, a newly
developed relativistic TD-CIS, RTDCIS that takes the Dirac equation as
its starting point \cite{Zapata2022}; Time-Dependent
Density-Functional Theory
\cite{Runge1984,Telnov1997,Telnov1998,DeGiovannini2012}; methods that
go beyond the single Slater determinant \emph{Ansatz}, such as
Multiconfigurational Time-Dependent Hartree
\cite{Meyer1990,Beck2000,Kvaal2011}, Multiconfigurational
Time-Dependent Hartree--Fock
\cite{Zanghellini2003,Zanghellini2004,Zanghellini2005,Nest2005,Hochstuhl2011},
Time-Dependent Multiconfigurational Self-Consistent-Field
\cite{Kato2004}; Time-Dependent Complete-Active-Space
Self-Consistent-Field \cite{Sato2013}; Time-Dependent Resolution in
Ionic States \cite{Spanner2009}; the various restricted active space
methods, e.g.\ Time-Dependent Restricted-Active-Space
Configuration-Interaction \cite{Hochstuhl2012} and Time-Dependent
Occupation-Restriction Multiple-Active-Space \cite{Sato2015}; the
excitation-class based methods Time-Dependent Coupled-Cluster
\cite{Huber2011,Kvaal2012,Sato2018a} and Algebraic Diagrammatic
Construction \cite{Ruberti2014,Ruberti2014a,Ruberti2019}; and finally
we mention the recent extension of the R-matrix method to molecules,
UKRmol+ \cite{Masin2020} and RMT \cite{Brown2020}. For reviews of the
various \emph{ab initio} approaches to multielectron dynamics, see
\textcite{Ishikawa2015,Armstrong2021}.

This article is arranged as follows: in section\ \ref{sec:equations}, the
general equations of motion for the TD-CIS \emph{Ansatz} are derived in
detail, as well as the generalization of surface-flux techniques to
compute photoelectron spectra to multiple ionization channels. The
time propagator is briefly surveyed in section\ \ref{sec:time-propagator},
and some illustrative calculations are presented in
section\ \ref{sec:results}. Finally, section\ \ref{sec:conclusions} concludes the
paper.

\subsection{Conventions}
\label{sec:org3d7695f}
\label{sec:conventions}
Hartree atomic units where \(\hbar=e=a_0=\electronmass=1\) are used
throughout.

We employ a modified version of Einstein's summation convention,
where indices appearing on one side of an equality sign only are
automatically contracted over, e.g.
\begin{equation*}
  \Psi = c_0\slater{0} + \contslater{k} \equiv
  c_0\slater{0} + \sum_k\contslater{k}.
\end{equation*}

For the canonical orbitals, we use the following letters:
\begin{itemize}
\item \(\ket{i}\), \(\ket{j}\), \(\ket{k}\), \(\ket{l}\) denote occupied orbitals,
\item \(\ket{a}\), \(\ket{b}\) denote virtual orbitals,
\item \(\ket{c}\), \(\ket{d}\), \(\ket{e}\), \(\ket{f}\) denote \emph{any} orbitals.
\end{itemize}
Matrix elements between orbitals are written using
Mulliken-like notation:
\begin{equation}
  \label{eqn:mulliken}
  \begin{aligned}
    \onebody{c}{d} &\defd
                     \matrixel{c}{\hamiltonian_0 + \laserinteraction(t)}{d}, \\
    \twobodydx{cd}{ef}
                   &\defd
                     \twobody{cd}{ef} - \twobody{cd}{fe}; \\
    \twobody{cd}{ef}
                   &\defd
                     \int
                     \frac{\diff{\spatialspin_1}\diff{\spatialspin_2}}{\abs{\vec{r}_1 - \vec{r}_2}}
                     \conj{\orbital{c}}(\spatialspin_1)
                     \conj{\orbital{d}}(\spatialspin_2)
                     \orbital{e}(\spatialspin_1)
                     \orbital{f}(\spatialspin_2),
  \end{aligned}
\end{equation}
where \(\hamiltonian_0\defd \kinop + \potop\) is the molecular
one-body Hamiltonian, and \(\laserinteraction(t)\) is the
time-dependent potential due to the external laser
field. \(\spatialspin_{1,2}\) refer to both spatial and spin coordinates of
the orbitals.

\section{General TD-CIS, equations and surface flux}
\label{sec:equations}
Our \emph{Ansatz} is
\begin{equation}
  \label{eqn:td-cis-ansatz}
  \Psi(t) = c_0(t)\slater{0} + \contslater{k}(t),
\end{equation}
where \(\slater{0}\) is the Slater determinant of the reference
state (typically the Hartree--Fock ground state), \(c_0(t)\) is
time-dependent complex amplitude, and \(\contslater{k}(t)\) an
excited Slater determinant obtained by substituting the occupied
orbital \(\ket{k}\) of the reference by a time-dependent particle
orbital \(\contket*{k}\) formed as a linear combination of the
virtual canonical orbitals:
\begin{equation*}
  \contslater{k}(t) \defd
  c_{ka}(t)
  \create{a}
  \annihilate{k}
  \slater{0}
  \implies
  \contket*{k} = c_{ka}(t)\ket{a}.
\end{equation*}
No distinction is made between excitation and ionization channels,
since \(\contket*{k}\) is associated with a particular hole
configuration of the remaining ion, the particle--hole density of
which is given by \(\ketbra*{\cont{k}}{k}\). We also note that this
implies that particle orbitals corresponding to different occupied
orbitals are non-orthogonal, \(\braket*{\cont{k}}{\cont{l}}\neq0\),
which must be taken into account when forming the energy expression
\cite{Loewdin1955a}.

To derive the equations of motion (EOMs), we start from the
Dirac--Frenkel variational principle:
\begin{equation}
  \label{eqn:dirac-frenkel}
  \vary{\bra{\Psi}}\Lagrangian = 0,
\end{equation}
where the Lagrangian is given by
\begin{equation}
  \label{eqn:td-cis-lagrangian}
  \Lagrangian =
  \matrixel{\Psi}{\Hamiltonian-\imdt}{\Psi} -
  \orthogonality{i}{\cont{j}} -
  \orthogonality{\cont{j}}{i}.
\end{equation}
The Lagrange multipliers \(\lagrange{i\cont{j}},\lagrange{\cont{j}i}\)
ensure that the particle orbitals \(\contket*{j}\) remain orthogonal
to the occupied orbitals \(\ket{i}\) at all times; this is necessary
since otherwise the basis would be overcomplete. In the time
propagation, they are implemented as projectors, without explicitly
evaluating the multipliers. A possible alternative approach would have
been to treat the Lagrange multipliers as dynamical variables
\cite{Car1985}. Such a formulation avoids an explicit
orthogonalization step, quadratic in the number of orbitals, leading
to substantial efficiency improvements in effective one-electron
theories such as TD-DFT. The potential savings are however likely to
remain minor in wavefunction-based methods such as TD-CIS.

Inserting the \emph{Ansatz} \eqref{eqn:td-cis-ansatz} into the expression
\eqref{eqn:td-cis-lagrangian} for the Lagrangian, we find
\begin{equation}
  \label{eqn:td-cis-lagrangian-inserted}
  \tag{\ref{eqn:td-cis-lagrangian}*}
  \Lagrangian =
  E -
  \conj{c_0}\imdt c_0 -
  \matrixel*{\cont{k}}{\imdt}{\cont{k}} -
  \orthogonality{i}{\cont{j}} -
  \orthogonality{\cont{j}}{i},
\end{equation}
where the total energy is given by
\begin{equation*}
  \begin{aligned}
    E
    &=
      \abs{c_0}^2 \refeng +
      \braket*{\cont{k}}{\cont{k}} E_k +
      \markterm{1}{(\conj{c_0}E_{k\tilde{k}} + \cc)} \\
    &\hphantom{=}+
      \markterm{2}{E_{\cont{k}\cont{k}}} -
      \braket*{\cont{k}}{\cont{l}}
      \markterm{3}{E_{lk}} -
      \markterm{4}{\twobodydx{l\cont{k}}{k\cont{l}}}.
  \end{aligned}
\end{equation*}
We have here introduced the following notation for the partial
contributions to the overall energy:
\begin{equation*}
  \refeng \defd \onebody{i}{i} + \frac{1}{2}\twobodydx{ij}{ij}
\end{equation*}
is the (time-dependent) energy of the reference determinant
\(\slater{0}\),
\begin{equation*}
  E_k \defd \refeng - \orbitalenergy{k}
\end{equation*}
the channel energy associated with excitation/ionization from the
occupied orbital \(k\), and the orbital energy
\begin{equation*}
  \orbitalenergy{k} \defd \orbitalenergy{kk},\quad
  \orbitalenergy{kl} \defd \onebody{k}{l} +
  \frac{1}{2}\twobodydx{kj}{lj}.
\end{equation*}
The occupied--virtual orbital energy (and its complex conjugate)
\begin{equation*}
  \markterm{1}{E_{k\cont{k}} \defd
    \onebody{k}{\cont{k}} +
    \twobodydx{ki}{\cont{k}i}}
\end{equation*}
contain the terms that lead to excitation/ionization from the
reference, whereas the virtual--virtual energy
\begin{equation*}
  \markterm{2}{E_{\cont{k}\cont{k}} \defd
    \onebody{\cont{k}}{\cont{k}} +
    \twobodydx{\cont{k}i}{\cont{k}i}}
\end{equation*}
describe the interaction of the excited/ionized electron with its
parent ion state (intrachannel) as well as with the external
field. Finally, the interchannel energies
\begin{equation*}
  \markterm{3}{E_{lk} \defd
    \onebody{l}{k} +
    \twobodydx{li}{ki}}; \quad
  \markterm{4}{\twobodydx{l\cont{k}}{k\cont{l}}}
\end{equation*}
describe coupling between the various ionization channels through
the external field and Coulomb interaction, respectively. We note
that due to the linearity of the \emph{Ansatz} \eqref{eqn:td-cis-ansatz},
we are free to choose the energy origin\footnote{This can be thought of
  as a gauge transform.}; by setting \(\refeng=0\), we avoid having to
converge the quick phase evolution due to the HF
reference. Simultaneously, the channel energy becomes simply
\(E_k=-\epsilon_k\).

By varying \eqref{eqn:td-cis-lagrangian-inserted} with respect to
\(\conj{c_0}\) and \(\contbra*{k}\), we get the EOMs:
\begin{equation}
  \label{eqn:td-cis-eoms}
  \begin{aligned}
    \imdt c_0
    &=
      c_0 \refeng +
      \markterm{1}{\matrixel*{k}{\fock}{\cont{k}}}, \\
    \imdt\contket*{k}
    &=
      (E_k+\markterm{2}{\fock})\contket*{k} +
      \markterm{1}{c_0\fock\ket{k}} \\
    &\hphantom{=}-
      \markterm{3}{\matrixel{l}{\fock}{k}}\contket*{l} -
      \markterm{4}{(\direct[lk]-\exchange[lk])\contket*{l}} -
      \lagrange{\cont{k}i}\ket{i},
  \end{aligned}
\end{equation}
where the Fock operator is defined as
\begin{equation*}
  \fock \defd
  \hamiltonian_0 +
  \laserinteraction(t) +
  \direct[ii] -
  \exchange[ii],
  \implies \matrixel{l}{\fock}{k}
  \equiv
  \orbitalenergy{lk},
\end{equation*}
and the direct and exchange potentials by their action on an
orbital:
\begin{equation*}
  \begin{aligned}
    \direct[cd]\ket{e}
    &\defd
      \orbital{e}(\spatialspin_1)
      \int\frac{\diff{\spatialspin_2}}{\abs{\vec{r}_1-\vec{r}_2}}
      \conj{\orbital{c}}(\spatialspin_2)
      \orbital{d}(\spatialspin_2), \\
    \exchange[cd]\ket{e}
    &\defd
      \orbital{d}(\spatialspin_1)
      \int\frac{\diff{\spatialspin_2}}{\abs{\vec{r}_1-\vec{r}_2}}
      \conj{\orbital{c}}(\spatialspin_2)
      \orbital{e}(\spatialspin_2)
      \equiv \direct[ce]\ket{d}.
  \end{aligned}
\end{equation*}
We now use our gauge freedom to choose \(\refeng=0\), and we get the
modified EOMs
\begin{equation}
  \label{eqn:td-cis-eoms-simplified}
  \tag{\ref{eqn:td-cis-eoms}*}
  \begin{aligned}
    \imdt c_0
    &=
      \markterm{1}{\matrixel*{k}{\fock}{\cont{k}}}, \\
    \imdt\contket*{k}
    &=
      (-\matrixel{k}{\fock}{k}+
      \markterm{2}{\fock})\contket*{k} +
      \markterm{1}{c_0\fock\ket{k}} \\
    &\hphantom{=}-
      \markterm{3}{\matrixel{l}{\fock}{k}}\contket*{l} -
      \markterm{4}{(\direct[lk]-\exchange[lk])\contket*{l}} -
      \lagrange{\cont{k}i}\ket{i},
  \end{aligned}
\end{equation}

If \(\ket{k}\) and \(\contket*{k}\) live in different vector spaces,
e.g.\ if \(\ket{k}\) is expanded in a basis set of Gaussians and
\(\contket*{k}\) is resolved on a grid, their respective matrix
representations of the Fock operator \(\fock\) will in general not
agree, i.e.\ \(\ket{k}\) resolved on the grid of \(\contket*{k}\) will
not necessarily be an eigenvector of the matrix representation of
\(\fock\) on the same grid. It is therefore important to compute
matrix elements of operators in the correct vector space. The chief
reason for representing the occupied orbitals using Gaussians instead
of resolving them on the grid and computing the matrix elements of the
Fock operator accordingly, is that any grid coarse enough to be
reasonable for time propagation would not be able to accurately
represent the oscillatory behaviour of the Fock operator close to the
nuclei. Specifically, the direct interaction \(\direct[ii]\) needs to
partially screen the nuclear potentials
\(-Z_B/\abs{\vec{R}_B - \vec{r}_i}\) such that the long range
potential behaves as \(-1/r_i\). This cancellation is challenging to
achieve accurately on a coarse grid, while it is trivial to compute
exactly in the Gaussian basis and then instantiate the result on the
grid. A similar argument can be made for the short-range, non-local
potential \(\direct[ii]-\exchange[ii]\).

If instead \(\ket{k}\) and \(\contket*{k}\) \emph{do} live in the same
vector space, we may choose \(\ket{k},\ket{l}...\) to be the canonical
orbitals which diagonalize the field-free Fock operator. This choice
(which we make in the special case of atomic symmetry; see the
following article) is a restriction compared to the more general case
considered in this article. Furthermore, in case of an atom placed at
the origin of our coordinate system, there are no permanent dipole
moments. These two restrictions considerably simplify the EOMs:
\begin{equation}
  \label{eqn:td-cis-eoms-atomic}
  \tag{\ref{eqn:td-cis-eoms}\(\dagger\)}
  \begin{aligned}
    \imdt c_0
    &=
      \markterm{1}{\matrixel*{k}{\laserinteraction}{\cont{k}}}, \\
    \imdt\contket*{k}
    &=
      (-\orbitalenergy{k}+
      \markterm{2}{\fock})\contket*{k} +
      \markterm{1}{c_0\laserinteraction\ket{k}} \\
    &\hphantom{=}-
      \markterm{3}{\matrixel{l}{\laserinteraction}{k}}\contket*{l} -
      \markterm{4}{(\direct[lk]-\exchange[lk])\contket*{l}} -
      \lagrange{\cont{k}i}\ket{i},
  \end{aligned}
\end{equation}
where we have dropped the term
\(\markterm{1}{c_0
  (\hamiltonian_0+\direct[ii]-\exchange[ii])\ket{k}}\) since we
require \(\braket*{k}{\cont{k}}=0\). Our molecular implementation
below uses \eqref{eqn:td-cis-eoms-simplified}, while our atomic
implementation \cite{Carlstroem2022tdcisII} relies on
\eqref{eqn:td-cis-eoms-atomic}.

\subsection{t+iSURF\{C,V\}}
\label{sec:orga9d6343}
The following derivation in similar in spirit to
\textcite{Scrinzi2012NJoP,Orimo2019}, but differs in the details.

We begin by defining the \(N\)-electron Heaviside function acting on
one electron:
\begin{equation}
  \label{eqn:asymptotic-heaviside}
  \Theta \defd
  \permute{jN}
  [1-\theta(r_1 - \matchradius)]
  ...
  [1-\theta(r_{N-1} - \matchradius)]
  \theta(r_N - \matchradius),
\end{equation}
where \(\permute{jN}\) permutes coordinates \(j\) and \(N\).
With this, we construct our \emph{Ansatz} for the asymptotic region
(\(r_j>\matchradius\)):
\begin{equation*}
  \begin{aligned}
    \Theta\ket{\Psi(t)}
    &=
      a_n(\vec{p}_\spin,t)
      \antisym
      \ket{\xi_n(t)}
      \ket{\vec{p}_\spin(t)} \\
    &\defd
      a_n(\vec{p}_\spin,t)
      \ket{\xi_n(t)\vec{p}_\spin(t)},
  \end{aligned}
\end{equation*}
where \(\{\ket{\xi_n(t)}\}\) is a \enquote{time-dependent, complete but
  otherwise arbitrary set of functions} \cite{Scrinzi2012NJoP} that
spans the ion degrees of freedom, and \(\ket{\vec{p}_\spin(t)}\) is
an electron with final momentum \(\vec{p}\) and spin projection
\(\spin\).

The antisymmetrization operator used above to couple an already
antisymmetrized (\(N-1\))-body wavefunction \(\ket{\xi_n(t)}\) with one
electron \(\ket{\vec{p}_\spin(t)}\) to form an antisymmetrized
\(N\)-body wavefunction is defined as
\begin{equation}
  \label{eqn:asymptotic-antisymmetrizer}
  \antisym \defd
  \frac{1}{\sqrt{N}}
  (-)^{N-j}
  \permute{jN},
\end{equation}
where \((-)^{N-j}\) is the signature of the permutation. This form is
valid, provided that the strong orthogonality assumption applies,
i.e.\ that the scattering state \(\ket{\vec{p}_\spin(t)}\) is
orthogonal to all the bound orbitals of the initial state
\cite{Martin1976,Pickup1977,Oehrn1981,Patchkovskii2006}. This
condition is trivially fulfilled due to the presence of the Heaviside
function \eqref{eqn:asymptotic-heaviside}.

The overlap with such an asymptotic state
is given by
\begin{equation}
  \label{eqn:scattering-state-overlap}
  \begin{aligned}
    a_n(\vec{p}_\spin,t)
    &=
      \matrixel{\xi_n(t)\vec{p}_\spin(t)}{\Theta}{\Psi(t)} \\
    &=
      \bra{\xi_n(t)}\!\!
      \matrixel{\vec{p}_\spin(t)}{\antisym\Theta}{\Psi(t)} \\
    &=
      \markterm{1}{\bra{\vec{p}_\spin(t)}}
      \theta\sqrt{N}
      \brakett{\markterm{3}{\xi_n(t)}}{\markterm{2}{\Psi(t)}} \\
    &\defd
      \matrixel{\markterm{1}{{\vec{p}_\spin(t)}}}{\theta}{n(t)},
  \end{aligned}
\end{equation}
where we have used the fact that \(\antisym\) is self-adjoint to act
with it on \(\Theta\ket{\Psi(t)}\) to the right, and that only \(N\) terms are
non-zero for which the coordinate of \(\ket{\vec{p}_\spin(t)}\)
coincides with the photoelectron orbital in \(\ket{\Psi(t)}\) (a single
orbital in TD-CIS). The sign of each such term will precisely
compensate the corresponding sign of the antisymmetrizer
\eqref{eqn:asymptotic-antisymmetrizer}. The normalization leaves a
factor of \(\sqrt{N}\), which is absorbed into the definition of the
Dyson orbital \(\ket{\dyson{n}(t)}\). Finally, \(\rAngle\) denotes
integration over both the photoelectron and ion coordinates. The
benefit of formulating the photoelectron amplitudes in terms of Dyson
orbitals is that the various surface terms then attain the familiar
expressions from the single-electron case, with the Dyson orbital
replacing the single-electron wavefunction. The only remaining
difference is in the channel coupling through the ion Hamiltonian.

It is convenient to choose \(\ket{\xi_n(t)}\) as the time-dependent
eigenstates of the ion:
\begin{equation*}
  \begin{aligned}
    \ket{\xi_n(t)}
    &\defd
      \ce^{-\im E_n (t-t_i)}
      \ket{\xi_n} \\
    \implies
    \bra{\xi_n(t)}
    [\imadjdt+E_n]
    &\equiv
      \bra{\xi_n(t)}
      [\imadjdt+\ionichamiltonian] = 0,
  \end{aligned}
\end{equation*}
where \(t_i\) is some reference time, usually taken as the
beginning of the laser interaction.

Similarly, the scattering states obey the one-electron TDSE
\begin{equation*}
  \bra{\vec{p}_\spin(t)}[\imadjdt+\scatteringhamiltonian] = 0
\end{equation*}
and the wavefunction in the asymptotic region
\begin{equation*}
  [\imdt-(\ionichamiltonian+\scatteringhamiltonian)]
  \ket{\Psi(t)} = 0.
\end{equation*}
The EOMs for the overlaps \eqref{eqn:scattering-state-overlap} are
then
\begin{widetext}
  \begin{equation*}
    \begin{aligned}
      -\imdt a_n(\vec{p}_\spin, t)
      &=
        \markterm{1}{
        \bra{\vec{p}_\spin(t)}\!
        \scatteringhamiltonian
        \theta
        \sqrt{N}\!
        \brakett{\xi_n(t)}{\Psi(t)}} +
        \markterm{3}{
        E_n
        \bra{\vec{p}_\spin(t)}\!
        \theta
        \sqrt{N}\!
        \brakett{\xi_n(t)}{\Psi(t)}} -
        \markterm{2}{
        \bra{\vec{p}_\spin(t)}\!
        \theta
        \sqrt{N}
        \matrixell*{\xi_n(t)}{\ionichamiltonian(t) +
        \scatteringhamiltonian}{\Psi(t)}
        } \\
      &=
        \markterm{3}{E_n a_n(\vec{p}_\spin, t)} -
        \markterm{2}{
        \bra{\vec{p}_\spin(t)}
        \theta
        \sqrt{N}\!
        \matrixell*{\xi_n(t)}{\ionichamiltonian(t)\identity}{\Psi(t)}
        } +
        \matrixel*{\vec{p}_\spin(t)}
        {
        \markterm{1}{\scatteringhamiltonian\theta}-
        \markterm{2}{\theta\scatteringhamiltonian}
        }
        {\dyson{n}(t)} \\
      &=
        \markterm{3}{E_n a_n(\vec{p}_\spin, t)} -
        \markterm{2}{
        \underbrace{\matrixel{\xi_n(t)}{\ionichamiltonian(t)}{\xi_m(t)}}_{\defd\mat{H}_{nm}(t)}
        \underbrace{\matrixel{\vec{p}_\spin(t)}{\theta}{\vec{q}_{\spin'}(t)}}_{\delta(\vec{p}-\vec{q})\delta_{\spin\spin'}}
        a_m(\vec{q}_{\spin'}, t)
        } +
        \underbrace{\matrixel*{\vec{p}_\spin(t)}
        {\comm*{\scatteringhamiltonian}{\theta}}
        {\dyson{n}(t)}}_{S_n(\vec{p}_\spin, t)} \\
      &=
        -\tilde{\mat{H}}_{nm}(t)
        a_m(\vec{p}_\spin, t) +
        S_n(\vec{p}_\spin, t),
    \end{aligned}
  \end{equation*}
\end{widetext}
where the ionic Hamiltonian in the interaction picture is given by
\begin{equation*}
  \begin{aligned}
    \tilde{\mat{H}}_{nm}(t)
    &\defd
      \mat{H}_{nm}(t) - \delta_{nm} E_n, \\
    \mat{H}_{nm}(t)
    &=
      \matrixel{\xi_n(t)}{\ionichamiltonian(t)}{\xi_m(t)} \\
    &=
      \ce^{\im E_{nm} t}[
      \delta_{nm}E_n +
      \matrixel{\xi_n}{\laserinteraction(t)}{\xi_m}
      ], \\
    E_{nm} &= E_n - E_m,
  \end{aligned}
\end{equation*}
and the asymptotic resolution of identity by
\begin{equation*}
  \identity =
  \ket{\xi_m(t)}
  \ketbra{\vec{q}_{\spin'}(t)}{\vec{q}_{\spin'}(t)}
  \bra{\xi_m(t)}.
\end{equation*}

The EOMs can be written on matrix form
\begin{equation}
  \label{eqn:tsurff-eoms}
  -\imdt \vec{a}(\vec{p}_\spin, t) =
  -\tilde{\mat{H}}(t)\vec{a}(\vec{p}_\spin, t) +
  \vec{S}(\vec{p}_\spin, t).
\end{equation}
\(S_n(\vec{p}_\spin,t)\) is the \emph{surface term} contributing to
ionization channel \(n\) and final photoelectron momentum
\(\vec{p}\) and spin projection \(\spin\). The precise form of the
surface terms depends on the asymptotic wavefunction chosen
(e.g.\ Coulomb or Volkov scattering wavefunctions), as well as the
gauge \cite{Tao2012NJoP,Morales2016-isurf}.

Rearranging, we get
\begin{equation}
  \tag{\ref{eqn:tsurff-eoms}*}
  \label{eqn:tsurff-eoms-rearranged}
  \begin{aligned}
    \partial_t \vec{a}(\vec{p}_\spin, t) +
    \im\tilde{\mat{H}}(t)\vec{a}(\vec{p}_\spin, t) =
    \im\vec{S}(\vec{p}_\spin, t),
  \end{aligned}
\end{equation}
which is an inhomogeneous differential equation for the
photoionization amplitude \(\vec{a}(\vec{p}_\spin, t)\). If we
discretize \eqref{eqn:tsurff-eoms-rearranged} in time and apply the
trapezoidal rule, evaluating the ionic Hamiltonian at half the time
step, we find (suppressing the dependence on \(\vec{p}_\spin\))
\begin{equation}
  \label{eqn:tsurff-solution}
  \begin{aligned}
    &
      \frac{\vec{a}_{i+1} - \vec{a}_i}{\timestep} +
      \frac{\im}{2}
      (\tilde{\mat{H}}_{i+1/2}
      \vec{a}_{i+1} +
      \tilde{\mat{H}}_{i+1/2}\vec{a}_i)
      =
      \frac{\im}{2}(
      \vec{S}_{i+1} +
      \vec{S}_i
      ) \\
    &\iff
      \left(\identity +
      \frac{\im\timestep}{2} \tilde{\mat{H}}_{i+1/2}
      \right)
      \vec{a}_{i+1}
      = \\
    &\qquad\qquad
      \left(\identity -
      \frac{\im\timestep}{2} \tilde{\mat{H}}_{i+1/2}
      \right)
      \vec{a}_i +
      \frac{\im\timestep}{2}
      (\vec{S}_i + \vec{S}_{i+1}), \\
  \end{aligned}
\end{equation}
which is globally accurate to \(\Ordo\{\timestep^2\}\).
\subsubsection{Diagonal \(\tilde{\mat{H}}\)}
\label{sec:org7303303}
If \(\tilde{\mat{H}}\) is diagonal (and time-independent), its
elements are zero, no channel-coupling occurs, and we can integrate
each component of \eqref{eqn:tsurff-eoms-rearranged} separately:
\begin{equation}
  \label{eqn:diagonal-tsurff-integrated}
  \begin{aligned}
    a_n(\vec{p}_\spin,t)
    &=
      \im
      \int_{t_i}^t\diff{\tau}
      S_n(\vec{p}_\spin,\tau) \\
    &=
      \im
      \int_{t_i}^t\diff{\tau}
      \bra{\vec{p}_\spin(\tau)}
      \comm*{\scatteringhamiltonian}{\theta}
      \ket{\dyson{n}(\tau)},
  \end{aligned}
\end{equation}
and we recover Koopman's theorem [since
\(\ket{\xi_n(\tau)}\equiv\ce^{-\im E_n(\tau-t_i)}\ket{\xi}\)],
which is an excellent approximation in CIS, if promotions are
allowed from the valence shell only.

\subsubsection{iSURF}
\label{sec:orgc0475d5}
In the case of a discretized spectrum \cite{Moiseyev2011}, the time
evolution of the wavefunction after the end of the pulse
(\(t=t_f\)) is trivially given by
\begin{equation}
  \label{eqn:decaying-wavefunction}
  \ket{\Psi(t>t_f)} =
  \ket{\gamma}
  \braket*{\gamma}{\Psi(t_f)}
  \ce^{-\im E_\gamma(t-t_f)}
\end{equation}
where \(\gamma\) represents all quantum numbers necessary to
describe an eigenstate of the field-free Hamiltonian
\(\Hamiltonian_0\), and \(E_\gamma\) is its eigenenergy, which
will have a negative imaginary part in the presence of absorbing
boundary conditions (corresponding to exponential decay of
outgoing waves). Furthermore, after the pulse has ended,
\(\tilde{\mat{H}}\) is diagonal, i.e.\ there is no coupling between
the ionization channels anymore, and only the surface term
remains, and the amplitude is given by
\eqref{eqn:diagonal-tsurff-integrated}. Integrating this amplitude
from the end of the pulse \(t_f\) to the detection time
(suppressing the \(\vec{p}_\spin\) argument), we find
\begin{equation*}
  a_n(\infty) - a_n(t_f) =
  \im
  \int_{t_f}^\infty\diff{\tau}
  \bra{\vec{p}_\spin(\tau)}
  \comm*{\scatteringhamiltonian}{\theta}
  \ket{\dyson{n}(\tau)}.
\end{equation*}
The total amplitude for channel \(n\) will be given by
\begin{equation*}
  \begin{aligned}
    a_n(\infty)
    &=
      a_n(t_f) +
      [a_n(\infty) - a_n(t_f)] \\
    &=
      \underbrace{a_n(t_f)}_{\textrm{tSURFF}}+
      \underbrace{\im\int_{t_f}^\infty\diff{\tau}S_n(\tau)}_{\textrm{iSURF}}.
  \end{aligned}
\end{equation*}

Inserting the \emph{Ansatz} \eqref{eqn:decaying-wavefunction} into the
source term, we find
\begin{equation*}
  \begin{aligned}
    &S_n(\vec{p}_\spin, t>t_f)
      =
      \bra{\vec{p}_\spin(t)}
      \comm*{\scatteringhamiltonian}{\theta}
      \ket{\dyson{n}(t)} \\
    &\quad
      =
      \bra{\vec{p}_\spin(t)}
      \comm*{\scatteringhamiltonian}{\theta}
      \sqrt{N}
      \braket{\xi_n(t)}{\gamma}
      \brakett{\gamma}{\Psi(t_f)}
      \ce^{\im(\epsilon - E_\gamma)(t-t_f)},
  \end{aligned}
\end{equation*}
where the \emph{target energy} is chosen as
\(\epsilon\defd E_n + p^2/2\). By identical manipulations as in
Equations~(14--17) of \textcite{Morales2016-isurf}, and assuming the
existence of the operator inverse, we arrive at
\begin{equation}
  \begin{aligned}
    \label{eqn:isurf}
    a_n(\infty) - a_n(t_f) =
    \matrixel{\vec{p}_\spin(t_f)}
    {\comm*{\scatteringhamiltonian}{\theta}}
    {\dyson{n}(\epsilon)},
  \end{aligned}
\end{equation}
where the Dyson orbital is given by
\begin{equation*}
  \ket{\dyson{n}(\epsilon)} =
  \sqrt{N}
  \braket{\xi_n(t_f)}{\Omega(\epsilon)},
  \quad
  \ket{\Omega(\epsilon)}\defd
  (\Hamiltonian_0-\epsilon)^{-1}\ket{\Psi(t_f)},
\end{equation*}
and \(\ket{\Omega(\epsilon)}\) is found using GMRES
\cite{Saad1986SJosasc}, with the \(\hamiltonian-\epsilon\) as the
preconditioner. We stress that Equations
\eqref{eqn:tsurff-eoms-rearranged}, \eqref{eqn:tsurff-solution},
and \eqref{eqn:isurf} are valid for single ionization for any
\emph{Ansatz}, and not just TD-CIS \eqref{eqn:td-cis-ansatz}.

\subsubsection{Surface terms, gauge dependence}
\label{sec:orgf1baa69}
The surface terms require the evaluation of
\begin{equation*}
  S_n(\vec{p}_\spin, t) =
  \matrixel*{\vec{p}_\spin(t)}
  {\comm*{\scatteringhamiltonian}{\heaviside}}
  {\dyson{n}(t)},
\end{equation*}
which depends on the particular form of scattering wavefunctions
chosen, as well as the gauge:
\begin{equation}
  \label{eqn:scattering-commutator}
  \begin{aligned}
    &
      \scatteringhamiltonian
      =
      \kinop + \laserinteraction(t) =
      \kinop +
      \begin{cases}
        \fieldamplitude[t]\cdot\vec{r}, & \textrm{(length gauge)},\\
        \vectorpotential[t]\cdot\vec{p}, & \textrm{(velocity gauge)},
      \end{cases} \\
    &\implies
      \comm*{\scatteringhamiltonian}{\heaviside}
      =
      \comm*{\kinop}{\heaviside} +
      \begin{cases}
        0, & \textrm{(length gauge)},\\
        \comm*{\vectorpotential[t]\cdot\vec{p}}{\heaviside}, & \textrm{(velocity gauge)}.
      \end{cases}
  \end{aligned}
\end{equation}

tSURFF has customarily been used in the velocity gauge, which is
valid since the SAE approximation is gauge invariant (provided
only local potentials are used)
\cite{Tao2012NJoP,Morales2016-isurf}. In this gauge, the Volkov
scattering wavefunctions have the simple expression
\begin{equation*}
  \volkovwave_{\vec{p}v}(t,\vec{r}) =
  \frac{1}{(2\pi)^{3/2}}
  \exp\left\{
    \im\vec{p}\cdot\vec{r}
    -\frac{\im}{2}
    \int_{t_0}^t\diff{t'}
    [\vec{p}+\vec{A}(t')]^2
  \right\},
\end{equation*}
the spatial part of which can be evaluated prior to time
propagation; only the Volkov phase varies with time.

If we instead work in the length gauge, we also need to gauge
transform the scattering wavefunctions according to
\begin{equation*}
  \begin{aligned}
    &\volkovwave_{\vec{p}l}(t,\vec{r})
      =
      \exp[\im\vec{A}(t)\cdot\vec{r}]
      \volkovwave_{\vec{p}v}(t,\vec{r}) \\
    &=
      \frac{1}{(2\pi)^{3/2}}
      \exp\left\{\im[\vec{p}+\vec{A}(t)]\cdot\vec{r}-
      \frac{\im}{2}
      \int_{t_0}^t\diff{t'}
      [\vec{p}+\vec{A}(t')]^2
      \right\},
  \end{aligned}
\end{equation*}
which means that when solving \eqref{eqn:tsurff-solution} in the
length gauge, the spatial part of the scattering wavefunctions need to
be reevaluated at every time step. However, as we see in Equation\
\eqref{eqn:scattering-commutator}, the laser coupling term in the
surface term vanishes.

We note that TD-CIS is \emph{not} gauge invariant
\cite{Wolfsberg1955,Kobe1979,Ishikawa2015}, and velocity gauge
calculations with fixed bound orbitals cannot be expected to agree
with length gauge calculations. Recently, \textcite{Sato2018}
introduced a \enquote{rotated velocity gauge} TD-CIS formulation, where
the gauge transform from length gauge to velocity gauge is applied
to the bound orbitals \enquote{on the fly}.

\begin{figure*}[tb]
  \centering
  \includegraphics{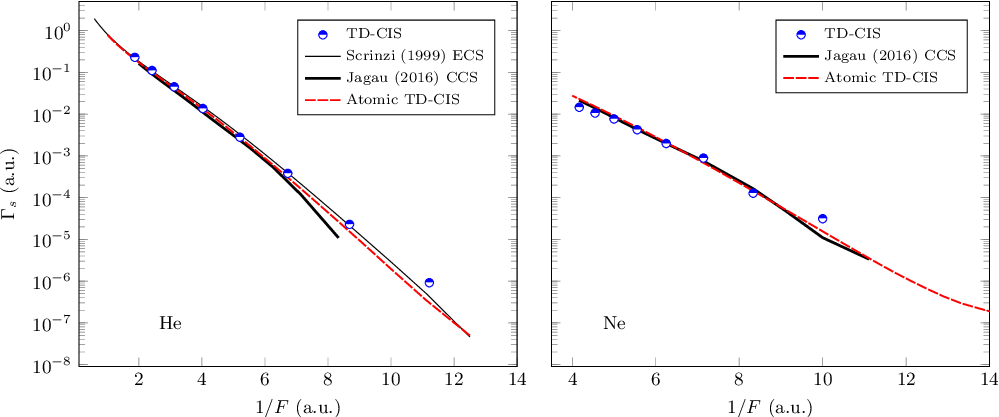}
  \caption{\label{fig:He_yields}Static-field ionization rates
    \(\Gamma_s\) for \helium{} and \neon{}. The rates computed with the
    current TD-CIS method are compared against the exact results of
    \textcite{Scrinzi1999} who used a full-dimensional complex scaling
    method, as well as \textcite{Jagau2016} who used a complex-scaled
    coupled-cluster approach. Also shown are the TD-CIS results
    obtained with the specialization to spherical symmetry (see
    following article).}
\end{figure*}

\section{Time propagator}

\label{sec:time-propagator} The EOMs
\eqref{eqn:td-cis-eoms-simplified} are solved by a simple
4\textsuperscript{th} order Runge--Kutta propagator, since that only
requires the action of the Hamiltonian on the wavefunction. The action
of the direct and exchange potentials on the particle orbitals
\(\contket*{k}\) is computed by solving Poisson's problem via
successive over-relaxation \cite{Young1950,Frankel1950}. The use of
iterative solvers means that the computational complexity scales
approximately linearly with the number of grid points, which in turn
is given by
\begin{equation*}
  n_g = \kappa n_xn_yn_zn_c, \quad
  \kappa =
  \begin{cases}
    1, & \textrm{spin-restricted},\\
    2, & \textrm{spin-unrestricted},
  \end{cases}
\end{equation*}
and \(n_c\) is the number of channels. Orthogonality of the particle
orbitals to the source orbitals is maintained by projecting out the
latter from the former, every time the Hamiltonian is applied.

To suppress reflections from the boundary of the computational domain,
we use the complex-absorbing potential (CAP) by
\textcite{Manolopoulos2002}, with a design parameter \(\delta=0.2\) leading
to \(<\SI{1}{\percent}\) reflection for all momenta above
\(k_{\textrm{min}}=\SI{1.5}{\au}\) The CAP then spans the last
\SI{4.37}{Bohr} of the box in each direction. To avoid lowering the
convergence order of the propagator, the CAP is multiplied by the time
step and applied as a mask function, separately from the other terms
in the Hamiltonian. The reason for this is that the convergence of the
propagator relies on the fact that the EOMs obey the Dirac--Frenkel
variational principle
\begin{equation}
  \label{eqn:dirac-frenkel-ortho}
  \tag{\ref{eqn:dirac-frenkel}*}
  \matrixel{\variation{\Psi}}{\Hamiltonian-\imdt}{\Psi} = 0
  \implies
  \bra{\variation{\Psi}}\perp\bra{\Psi}.
\end{equation}
If the EOMs are derived assuming a Hermitian \(\Hamiltonian\), the
presence of the CAP may result in the EOMs no longer obeying
\eqref{eqn:dirac-frenkel-ortho}. Applying the CAP as a mask function
circumvents the issue.

\section{Results}
\label{sec:results}
We now outline the capabilities of our TD-CIS implementation using
\helium{}, \neon, and a variety of small molecules: \lih{}, \water{},
and \ethylene{}. In all cases the GAMESS-US \cite{Schmidt1993}
electronic structure program was used to compute the initial
restricted Hartree--Fock (RHF) molecular orbitals (MOs); since only the
basis set and the corresponding expansion coefficients of the MOs are
required as inputs for the TD-CIS computations, any electronic
structure program may be used; all the required one- and two-electron
matrix elements are then evaluated by a Gaussian integral package
internal to the TD-CIS code. The \basisset{aug-cc-pVTZ} basis set was
used for each system.

Except for the case of \helium{}, the TD-CIS computations were all
carried out using continuum grids that extended from
\SIrange{-22.5}{+22.5}{Bohr} in all spatial directions with
\(\num{270}\times\num{270}\times\num{270}\) grid points, yielding a spatial step
size of \(\spatialstep\approx\SI{0.167}{Bohr}\). The time step in all cases
was \(\timestep=\SI{0.002}{\au}\approx\SI{0.05}{\atto\second}\).

\subsection{Atoms; helium and neon}

In the case of helium, accurate static-field ionization rates are
available in the literature. \textcite{Scrinzi1999}, in particular,
computed the \helium{} ionization rates using a full-dimensional
treatment within the exterior complex scaling (ECS)
methodology. \textcite{Jagau2016} used a complex-scaled
coupled-cluster approach, at various levels of theory for the
field-free reference state. We use these rates as a benchmark to
confirm the validity of our TD-CIS implementation. The spatial grid
extends from \SIrange{-15}{+15}{Bohr}, in all spatial directions with
\(\num{180}\times\num{180}\times\num{180}\) grid points, yielding a spatial step
size \(\spatialstep\approx\SI{0.168}{Bohr}\).

In order to compute the static-field ionization rates within
TD-CIS, a time-dependent computation is run with the system
initialized in the neutral state.  With a static field applied, the
time-dependent population of the neutral, \(\abs{c_0(t)}^2\), is
monitored. After an initial turn-on transient has passed, the
long-time behaviour of \(\abs{c_0(t)}^2\) is fit to an exponential
decay (\(\sim e^{-\Gamma_s t}\)) to extract the static-field
ionization rate \(\Gamma_s\).  Figure~\ref{fig:He_yields} shows the TD-CIS
ionization rates, which are in excellent agreement with those
obtained by \textcite{Scrinzi1999,Jagau2016}. At lower intensities, our
computational box is not sufficiently large enough to contain the
tunnel exit (see the discussion for \water{} below). The apparent
ionization rates at these intensities are the upper bound to the
true tunnelling rate.

\subsection{Convergence}
\begin{figure}[tb]
  \centering
  \includegraphics{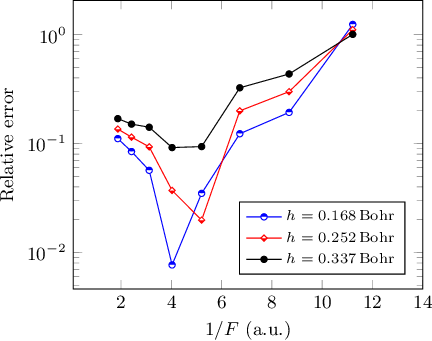}
  \caption{Relative errors of the \helium{} static field ionization
    rates shown in Figure~\ref{fig:He_yields} with respect to the
    atomic TD-CIS results, for various grid spacings.}
  \label{fig:static-rates-convergence}
\end{figure}

As a test of the convergence of the calculations, we compute the
static ionization rates for \helium{} for different grid spacings and
time steps. We compare with the results from the atomic TD-CIS
implementation, which is at the same level of theory and thus serves
as our method limit. In figure~\ref{fig:static-rates-convergence}, the
convergence with respect to the grid spacing is shown. Its behaviour
is non-uniform, reflecting the underlying complexity of the iterative
algorithms. For the different time steps tried, i.e.
\(\timestep\in\{0.001,0.002,0.004,0.008,0.016\}\;\si{\au}\), the change
in the error was negligible for all field strengths \(F\) and grid
spacings \(\spatialstep\). The convergence with respect to the time
step is most likely limited by the tolerance set for the Poisson
solver employed for the Coulomb interaction.

\subsection{\lih{}}
\begin{figure}[t]
  \centering
  \includegraphics{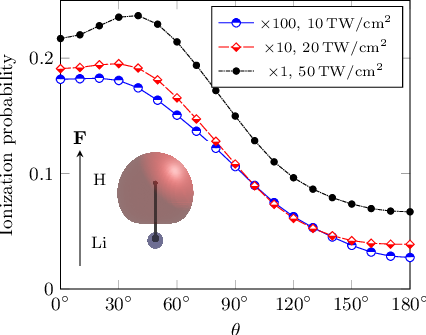}
  \caption{\label{fig:LiH_yields} Angle-dependent half-cycle
    ionization yields for \lih{} for intensities of
    \SI{10}{\tera\watt\per\centi\meter\squared} (data scaled by
    \(\times\)100), \SI{20}{\tera\watt\per\centi\meter\squared} (data
    scaled by \(\times\)10), and
    \SI{50}{\tera\watt\per\centi\meter\squared}. Here \(\theta\) is the
    angle between the molecular axis and the electric field vector of
    the laser. We clearly see that doubling the intensity
    (\(\sim F_0^2\)) increased the ionization probability by more than
    one order of magnitude. The inset shows the ionizing orbital,
    along with the electric field vector \(\vec{F}\) for the
    \(\theta=\ang{0}\) configuration where the field points from the
    \ch{Li} to the \ch{H} atom.}
\end{figure}
The \lih{} molecule has two occupied orbitals in the RHF neutral
ground state.  However, due to the large binding potential of the
\HOMO[-1] orbital, only the \HOMO{} (highest occupied molecular
orbital) orbital ionizes with non-negligible probability. Hence,
we treat LiH as a single channel case, with the lower-lying RHF
orbital frozen during the computations.

We calculate the half-cycle ionization yield using a smoothed
half-cycle pulse defined as
\begin{equation}
  \label{eqn:HalfCycle}
  F(t) =
  \begin{cases}
    0, & t < 0, \\
    F_0 \sin^2(\alpha t), & 0\leq t\leq \tmax, \\
    0, & t > \tmax,
  \end{cases}
  \quad
  \alpha\defd\frac{\omega_L}{\sqrt{2}},
\end{equation}
where \(\tmax=\pi/\alpha\). This pulse shape mimics the
high-intensity portion of a half-cycle of a laser field with
frequency \(\omega_L\), but using the smoothed half-cycle reduces
artefacts that arise from an instantaneous turn-on had just a
standard half-cycle field been used instead. For all remaining
computations, we use \(\omega_L=\SI{0.057}{\hartree}\), which
corresponds to an \SI{800}{\nano\meter} laser field. In addition to
minimizing the computational load required to compute the
ionization compared to that needed for a multi-cycle pulse, using a
half-cycle-type pulse allows us to see the orientation-dependence
of the ionization yields when ionizing polar molecules like \lih{}
(and \water{} considered below).

\begin{figure}[t]
  \centering
  \includegraphics{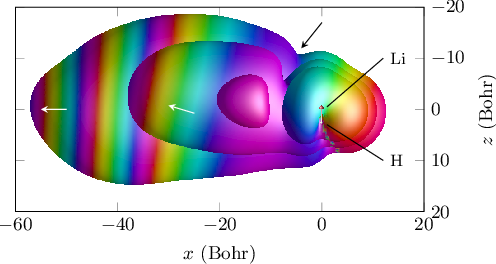}
  \includegraphics{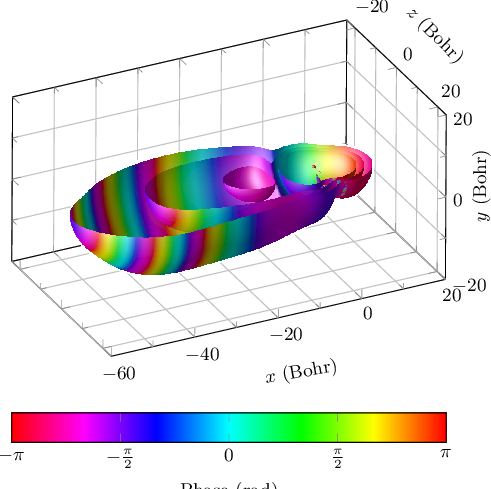}
  \caption{\label{fig:LiH_yields:EWP} Continuum electron wave packet
    after the end of the pulse [\(t=\tmax\) in \eqref{eqn:HalfCycle}],
    for \(\theta=\ang{90}\), and
    \SI{20}{\tera\watt\per\centi\meter\squared}.}
\end{figure}

The angle dependent half-cycle ionization yields are presented in
Figure~\ref{fig:LiH_yields} for the laser intensities for a variety of
laser intensities.  The inset shows the definition of the angle
between the molecular axis and the electric field of the ionizing
pulse; \(\theta=\ang{0}\) corresponds to the electric field vector of the
laser pointing from the Li to the H atom. Since the motion of the
liberated electron will be opposite to the direction of the electric
field, the angle corresponding to the peak of the angle-dependent
ionization (\(\theta \approx \ang{0}-\ang{45}\)) sees the liberated electron
escape from the initial \HOMO{} orbital (located on the \ch{H} atom)
across the \ch{Li} atom.

Figure~\ref{fig:LiH_yields:EWP} shows the continuum electron wave
packet at the end of the laser pulse, for the case of
\(\theta=\ang{90}\), i.e\ the laser field is perpendicular to the
molecular axis. The tunnel barrier is clearly visible around
\(x\approx\SI{-5}{Bohr}\), in that the wave packet appears \enquote{pinched}
(marked with a black arrow). The colour map which corresponds to the
phase indicates that the wavefront (and hence the direction of travel)
right after the tunnel exit is directed upwards (smaller value of
\(\theta\)), but after some distance, the electron motion tends to
\(\ang{-90}\), i.e\ parallel to \(-x\), opposite to the field
polarization (marked with white arrows). This agrees with the
observation above that the electron preferentially escapes the
potential in the vicinity of the \ch{Li} atom.


\subsection{\water{}}

\begin{figure}[tb]
  \centering
  \includegraphics{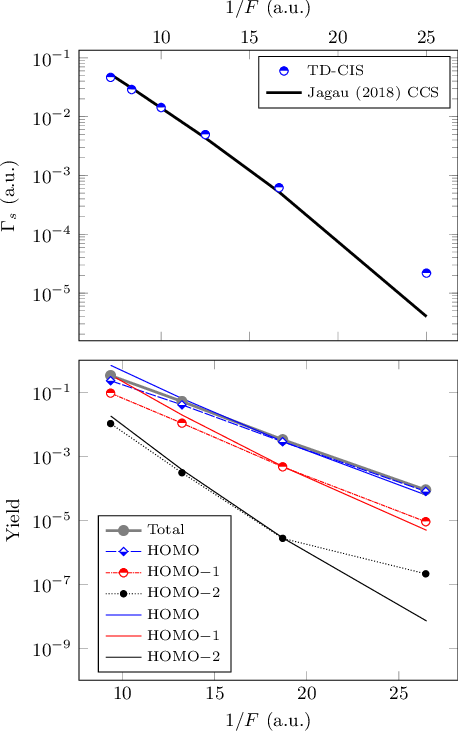}%
  \caption{\label{fig:Water_yields} \emph{Top panel}: Static
    ionization rates of \water{}, orthogonal to the molecular plane,
    compared with the CCS results of \textcite{Jagau2018}. \emph{Bottom
      panel}: Half-cycle intensity-dependent photoioniziation yield of
    \water{}, total, as well as resolved on the RHF orbitals. The
    \enquote{knee} observed at lower intensities for \HOMO[-2] is a
    polarization effect (see main text).  For comparison, the solid
    lines indicate the ionization yields as predicted by an
    Keldysh-like theory \eqref{eqn:keldysh-rate}, where only the
    exponential factor due to the ionization potential of the
    molecular orbitals, and the field strength is considered. These
    lines have been normalized to the respective TD-CIS results at
    \SI{100}{\tera\watt\per\square\centi\meter}.}
\end{figure}

Due to the lower symmetry of \water{}, belonging to the
\pointgroup{C}{2\vrefl} point group, it is not enough to compute the
photoionization yields as a function of \(\theta\) only. We therefore
choose a 9th-order Lebedev grid \cite{Lebedev1975} that can represent
integrals involving spherical harmonics up to
\(\ell=\left\lfloor\frac{9-1}{2}\right\rfloor=4\) exactly. From this grid of 38 field
orientations, only 16 that are unique with respect to the
\pointgroup{C}{2\vrefl} symmetry are actually computed, and then
replicated. The resulting yield surface is then transformed to the
equivalent expansion in spherical harmonics; the results are shown in
Figure~\ref{fig:Water_yields}. For the three highest occupied
molecular orbitals and four different intensities of the driving
field, the ionization yields are plotted as a function of the angles
\(\theta\) and \(\phi\), as well as angle-integrated. The same field
\eqref{eqn:HalfCycle} is used as for \lih, i.e\ a half-cycle of
\SI{800}{\nano\meter}. Additionally, static ionization rates are
computed (similarly to \helium{} and \neon) orthogonal to the
molecular plane, i.e\ along the lobes of the \HOMO{} (the \(x\) axis),
and compared with the CCS results of \textcite{Jagau2018}.

The dynamic photoionization yields can be compared to tunnelling
yields estimated from a Keldysh-like \cite{Keldysh1965} formula
for the ionization rate:
\begin{equation}
  \label{eqn:keldysh-rate}
  \begin{aligned}
    w
    &\sim
      \exp\left[-\frac{1}{3\tilde{\mathcal{E}}}g(\gamma)\right], \\
    g(\gamma)
    &\defd
      \frac{3}{2\gamma}\left[ \left(1 +
      \frac{1}{2\gamma^2}\right)\arcsinh\gamma - \frac{\sqrt{1 +
      \gamma^2}}{2\gamma} \right],
  \end{aligned}
\end{equation}
where \(\gamma\defd \sqrt{\ionpotential/2\ponderomotive}\),
\(\tilde{\mathcal{E}}\defd F[2(2\ionpotential)^{3/2}]^{-1}\), and
the estimated ionization yield \(y=w\tmax\). We attribute the
deviations of the predictions by the tunnelling formula from the
TD-CIS results to the neglect of many-electron dynamics in the
molecule in \eqref{eqn:keldysh-rate}.

\begin{table}[htbp]
  \caption{\label{tab:water-tunnel-exits}Tunnel exit
    \eqref{eqn:classical-tunnel-exit} for \water{}; missing values
    indicate barrier suppression. The MO energies are given in the
    second row. The CAP starts at \SI{18.13}{Bohr}.}
  \centering
  \myruledtabular{S[table-format=3]S[table-format=1.3]|S[table-format=2.3]S[table-format=2.3]S[table-format=2.3]}{%
    \mc{Intensity}\Tstrut & \mc{\(F\)} & \mc{\HOMO} & \mc{\HOMO[-1]} & \mc{\HOMO[-2]}\\
    \mc{(\si{\tera\watt\per\centi\meter\squared})}\Bstrut & \mc{(\si{\au})} & \SI{0.511}{Ha} & \SI{0.585}{Ha} & \SI{0.723}{Ha} \\
    \hline
    400\Tstrut & 0.107 &  &  & \\
    200 & 0.075 &  &  & 6.733\\
    100 & 0.053 & 6.727 & 8.328 & 11.106\\
    50\Bstrut & 0.038 & 11.100 & 13.141 & 16.887
  }
\end{table}
The non-exponential behaviour of the \HOMO[-2] yield for the
smallest intensity is an artefact of limited computational box
size. We compute the classical tunnel exit according to
\begin{equation}
  \label{eqn:classical-tunnel-exit}
  \rexit
  = \frac{2\ionpotential}{4F} +
  \sqrt{\left(\frac{2\ionpotential}{4F}\right)^2-\frac{2\ionpotential}{F}},
\end{equation}
where \(\ionpotential\) is the magnitude of the orbital energy in
Koopman's approximation, and \(F=\sqrt{I}\) the peak amplitude of the
ionizing field. Table~\ref{tab:water-tunnel-exits} shows the estimated
tunnel exits for the three highest occupied molecular orbitals and the
various field strengths. For the weakest field, we see that the tunnel
exit of \HOMO[-2] is very close to the onset of the CAP. This means
that in that particular channel, Rydberg states (which extend into the
classically forbidden region in the tunnel but not beyond it)
incorrectly count towards the photoionization yield, since we compute
ionization as loss of norm. Hence, what we observe is rather a
polarization effect, than true ionization, for these MOs. Also the
angular distribution of the yield for \HOMO[-2] is affected by this
artefact, resulting in a qualitatively different shape for the lowest
intensity. This error decreases with increased box size, at the cost
of longer computation times.
\begin{figure}[htb]
  \centering
  \includegraphics{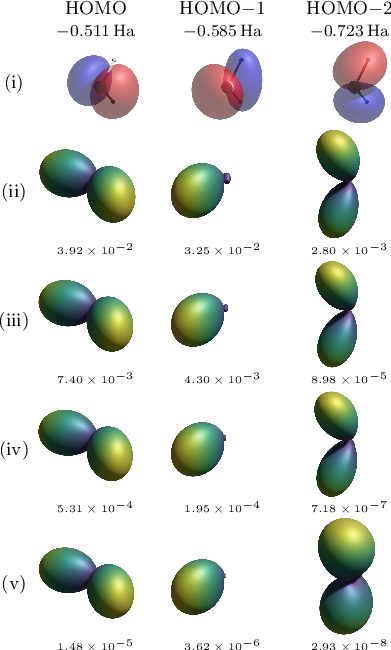}
  \caption{\label{fig:h2o-yields} (i) RHF orbitals of \water{}, along
    with their angle-resolved ionization yields, for the following
    intensities: (ii) \SI{400}{\tera\watt\per\centi\meter\squared},
    (iii) \SI{200}{\tera\watt\per\centi\meter\squared}, (iv)
    \SI{100}{\tera\watt\per\centi\meter\squared}, (v)
    \SI{50}{\tera\watt\per\centi\meter\squared}.  The yields have been
    normalized to their respective maximum values, indicated below
    each distribution.}
\end{figure}

\begin{table}[htbp]
  \caption{\label{tab:water-mo-overlaps}Overlaps between the molecular orbitals of the \water{} ground state and those of the \ch{O} ground state. We see that to a very large degree, the MOs of \water{} are atomistic.}
  \centering
  \myruledtabular{l|S[table-format=1.3]S[table-format=1.3]S[table-format=1.3]}{%
    & \mc{\ch{O} \(\conf{p}_x\)}\Tstrut\Bstrut & \mc{\ch{O} \(\conf{p}_y\)} & \mc{\ch{O} \(\conf{p}_z\)}\\
    \hline
    \ch{H2O} \HOMO\Tstrut & 0.995 & 0.000 & 0.000\\
    \ch{H2O} \HOMO[-1] & 0.000 & 0.000 & 0.951\\
    \ch{H2O} \HOMO[-2] & 0.000 & 0.944 & 0.000
  }
\end{table}
We now turn to the angular distributions of the photoionization
yields; as we can see in the first row of
Figure~\ref{fig:h2o-yields}, the molecular orbitals (MOs) are very
similar to atomic orbitals (AOs) of \conf{p} symmetry. The small
deviation of the MOs from pure spherical symmetry leads however to
drastically different angularly resolved ionization yields, where
the electron preferentially leaves the molecule over the \ch{O}
atom, which we attribute primarily to dipole effects. This pattern
persists for all intensities, although the \HOMO[-1] contribution
in the direction of the two \ch{H} atoms does seem to increase
slightly with intensity [middle column of row (ii) in
Figure~\ref{fig:h2o-yields}]. The nodal planes in the \HOMO{} and
\HOMO[-2] yields are mandated by the \pointgroup{C}{2\vrefl}
symmetry. The observed nodal plane in the \HOMO[-1] yields is
however required by an approximate, higher symmetry,
\pointgroup{D}{2\hrefl}. This serves to further highlight the
atomistic nature of the \ch{H2O} MOs, which we can quantitatively
investigate by computing their overlaps with the \conf{p} AOs of
\ch{O}; see Table~\ref{tab:water-mo-overlaps}.

\begin{figure}[htb]
  \centering
  \includegraphics{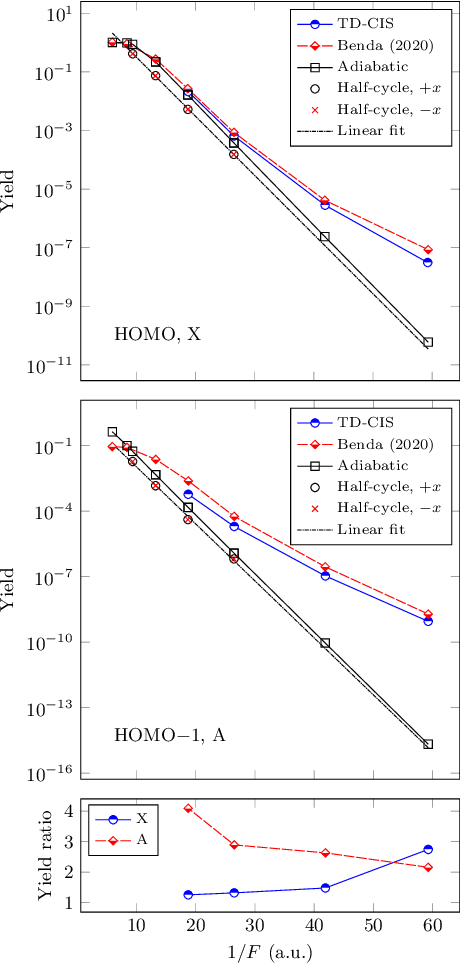}
  \caption{\label{fig:Water_dynamic_yields}\water{} yields for
    ionization from the two highest-lying occupied orbitals using an
    eight-cycle pulse, as a function of inverse field strength. The
    yields are compared with the results of \textcite{Benda2020}, as
    well as with an adiabatic estimate \eqref{eqn:adiabatic-yield}
    formed from the half-cycle yields presented in
    Figure~\ref{fig:Water_yields}. The bottom panel shows the ratio
    between the results of \textcite{Benda2020} and the present work.}
\end{figure}
In Figure~\ref{fig:Water_dynamic_yields}, the photoionization yields
along \(x\) for an eight-cycle pulse with a \(\sin^2\) envelope, and a
carrier wavelength of \SI{800}{\nano\meter} are shown. The yields are
compared with the results of \textcite{Benda2020} (specifically their
coupled model B), and show very good agreement. The discrepancy can
be mostly attributed to the difference in ionization potentials; in
the present work, the MOs are slightly more bound than in the
calculations by \textcite{Benda2020} (\SI{13.90}{\electronvolt} vs.\
\SI{13.15}{\electronvolt} for \HOMO{}, and \SI{15.92}{\electronvolt}
vs.\ \SI{15.15}{\electronvolt} for \HOMO[-1]), which has a strong
effect due to the exponential sensitivity of strong-field ionization
to the ionization potential.

Also shown are the half-cycle yields \(y_n\) in the lower panel of
Figure~\ref{fig:Water_yields}, along with their logarithms
\(\log_{10}(y_i)\) appear to follow a linear trend as a function of
the inverse field strength \(1/F_n\). We may therefore estimate the
eight-cycle yield adiabatically as
\begin{equation}
  \label{eqn:adiabatic-yield}
  \begin{aligned}
    y
    &=
      1 - \exp\left\{-\int_{-\infty}^\infty\diff{t} \Gamma[F(t)]\right\} \\
    &\approx
      1 - \exp\left[-\sum_i \expect{\Gamma(F_i)}\right] =
      1 - \prod_i [1-\tilde{y}(F_i)],
  \end{aligned}
\end{equation}
where \(F_i\) is the peak field amplitude of the \(i\):th
half-cycle, and the corresponding yields \(\tilde{y}(F_i)\) are
found using an linear fit to the half-cycle yields \(y_n\). This
estimate of course neglects any cycle-to-cycle effects, which
explains its deviation from the true eight-cycle results at lower
intensities (higher \(1/F\)). For higher intensities, the adiabatic
results are however in good agreement with the full calculations.

\subsection{\ethylene{}}
\begin{figure}[b]
  \centering
  \includegraphics{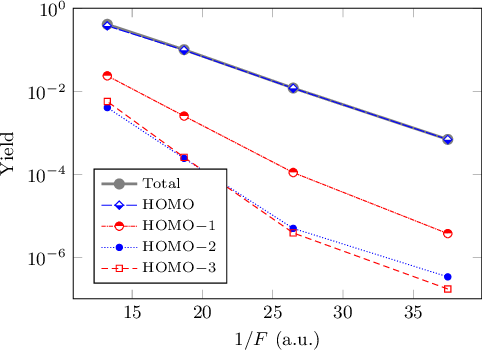}%
  \caption{\label{fig:c2h4_yields}Intensity-dependent photoioniziation
    yield of \ethylene{}, total as well as resolved on the RHF
    orbitals.}
\end{figure}
\begin{figure}
  \centering
  \includegraphics{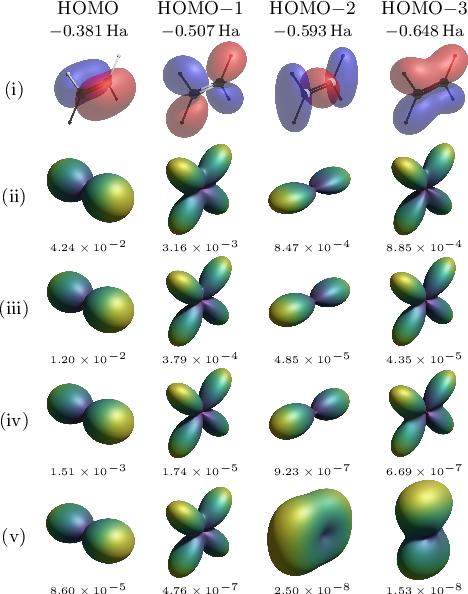}
  \caption{\label{fig:c2h4-angular-yields} (i) RHF orbitals of \ethylene{},
    along with their angle-resolved ionization yields, for the following
    intensities: (ii) \SI{200}{\tera\watt\per\centi\meter\squared},
    (iii) \SI{100}{\tera\watt\per\centi\meter\squared}, (iv)
    \SI{50}{\tera\watt\per\centi\meter\squared}, (v)
    \SI{25}{\tera\watt\per\centi\meter\squared}.  The yields have been
    normalized to their respective maximum values, indicated below each
    distribution.}
\end{figure}

Ionization yields are computed similarly to \water, as detailed
above, but \ethylene{} belonging to the \pointgroup{D}{2\hrefl}
point group, has higher symmetry than \water, and only 9 unique
field orientations are required. The results are shown in
Figure \ref{fig:c2h4_yields}, and the corresponding angular
distributions in Figure \ref{fig:c2h4-angular-yields}.

\begin{table}[h]
  \caption{\label{tab:ethylene-tunnel-exits}Tunnel exit
    \eqref{eqn:classical-tunnel-exit} for \ethylene{}; missing values
    indicate barrier suppression. The MO energies are given in the
    second row. The CAP starts at \SI{18.13}{Bohr}.}
  \centering
  \myruledtabular{S[table-format=7]S[table-format=2.3]|S[table-format=2.4]S[table-format=2.6]S[table-format=2.6]S[table-format=2.6]}{%
  \mc{Intensity}\Tstrut & \mc{\(F\)} & \mc{\HOMO} & \mc{\HOMO[-1]} & \mc{\HOMO[-2]} & \mc{\HOMO[-3]}\\
  \mc{(\si{\tera\watt\per\centi\meter\squared})}\Bstrut & \mc{(\si{\au})} &
  \SI{0.381}{Ha} &
  \SI{0.507}{Ha} &
  \SI{0.593}{Ha} &
  \SI{0.648}{Ha} \\
  \hline
  200\Tstrut & 0.075 &  &  &  & 5.412\\
  100 & 0.053 &  & 6.636 & 8.494 & 9.615\\
  50 & 0.038 & 7.347 & 10.988 & 13.360 & 14.858\\
  25 & 0.027 & 11.871 & 16.726 & 19.997 & 22.081
}
\end{table}

\begin{figure}
  \centering
  \includegraphics{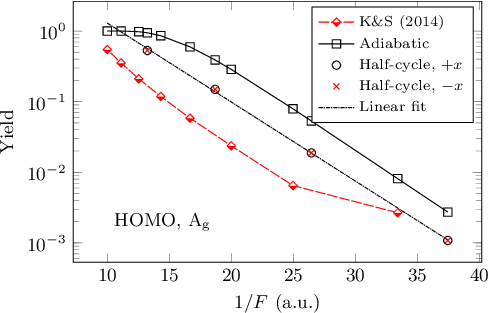}
  \caption{\label{fig:c2h4_dynamic_yields}\ethylene{} adiabatic
    dynamic yields \eqref{eqn:adiabatic-yield} for a seven-cycle
    pulse, compared with the results of \textcite{Krause2014a}.}
\end{figure}
Again, as in the case of \water, we find that for the lower
intensities, the yields for \HOMO[-\{2,3\}] deviate from the expected
exponential behaviour, with a \enquote{knee} observed in the
integrated ionization yield, and qualitatively different angular
distributions [row (v) in Figure \ref{fig:c2h4-angular-yields}],
whereas \HOMO{} and \HOMO[-1] follow their respective trends also for
the lowest intensity. Table \ref{tab:ethylene-tunnel-exits} lists the
tunnel exits for the various MOs and intensities, and for
\HOMO[-\{2,3\}] and the lowest intensity, these are well beyond the
CAP onset, implying we are measuring mainly a polarization effect.

In Figure \ref{fig:c2h4_dynamic_yields} we compare an adiabatic
estimation according to \eqref{eqn:adiabatic-yield} of the seven-cycle
yield with the results by \textcite{Krause2014a}. Their quoted
ionization rates were multiplied by the duration of the pulse [cf.\
their equation (8)] to obtain the yields; we do not find agreement
between the different calculations. Extracting the true ionization
rates is non-trivial, due to the temporal intensity averaging, which
is why we cannot easily compare directly with the half-cycle yields
presented in Figure \ref{fig:c2h4_yields}. We also note the
experimental work of \textcite{Talebpour1999}, the comparison with
which would require detailed knowledge of the experimental conditions,
not available to us, as well as extensive simulations.

\section{Conclusions}
\label{sec:conclusions}
We have described an implementation of TD-CIS for small gas-phase
molecules, with the continuum electron resolved on a Cartesian grid,
which allows an accurate description of the photoelectron in
strong-field processes. We applied this method to a few molecules,
finding reasonable results for the angularly resolved ionization
probability. An implementation of t+iSURF for TD-CIS in the atomic
case is presented in the following article
\cite{Carlstroem2022tdcisII}, and for the general case an
implementation is underway, which would allow us to efficiently
compute photoelectron spectra for various processes of interest.

\begin{acknowledgments}
  The work of SCM has been supported through scholarship 185-608 from
  \emph{Olle Engkvists Stiftelse}.
\end{acknowledgments}

\bibliography{tdcis-1-molecules}


\end{document}